# Magnetic Order in Pulsed Laser Deposited $(Fe,Ni)_5GeTe_2$ Films

Tamal Kumar Dalui[1], John Derek Demaree[2], Thomas Parker[2] and Ramesh C. Budhani[1,*]

[1]*Department of Physics and Engineering Physics, Morgan State University, Baltimore, Maryland-21251, USA.*

[2]*DEVCOM Army Research Laboratory, Aberdeen Proving Ground, Aberdeen, MD 21005, USA.*

**Corresponding author email: ramesh.budhani@morgan.edu*

## Abstract

We report the successful growth of highly textured thin films of $(Fe,Ni)_5GeTe_2$ two-dimensional ferromagnet on c-plane sapphire using pulsed laser deposition. Structural characterization via X-ray diffraction confirms preferential orientation along the (000l) direction, indicative of a high crystallographic texture. These films of van der Waals (vdW) type interplanar bonding exhibit robust ferromagnetism with a Curie temperature reaching ≈ 498 K. Electrical transport measurements reveal a clear anomalous Hall effect, with an anomalous Hall conductivity and Hall angle (%) of ≈ 20 $\Omega^{-1}cm^{-1}$ and ≈ 0.90, respectively. Furthermore, the magnetoresistance displays a pronounced dependence on film thickness, highlighting the tunability of spin-dependent transport in these vdW ferromagnetic thin films.

Two-dimensional (2D) van der Waals (vdW) magnets have emerged as an exciting class of materials to study long-range ordering of atomic moments in the low-dimensional limit [1–6]. The vdW materials such as $Cr_2Ge_2Te_6$, $CrI_3$, $CrTe_2$, $Fe_NGeTe_2$ (N = 3, 4 & 5) etc have been extensively investigated as model systems to explore quasi-2D magnetism [7–14]. Amongst these, $Fe_5GeTe_2$ (F5GT) has garnered particular interest owing to the rich spectrum of intriguing magnetic phenomena that it displays, including the emergence of magnetic skyrmions, a large anomalous Hall effect, planar topological Hall effect, and giant tunnelling magnetoresistance [15–19]. Notably, F5GT exhibits a Curie temperature ($T_C$) exceeding ~ 300 K, significantly higher than that of F3GT ($T_C$ ~150 –230 K), making it especially promising for practical applications. Recent investigations have underscored the critical role of Fe vacancies and the presence of multiple Fe sublattices in governing the complex magnetic behavior of F5GT [20–22]. However, the application of vdW magnets at ambient temperatures demands new compounds with $T_C$ well above the $T_C$ of F5GT. Various experimental strategies have been explored to enhance the $T_C$ of the F5GT compounds, including elemental doping, electrostatic gating, and strain engineering [6,23,24]. Moreover, for spintronic technologies, it is also desirable to fine-tune experimental methods for large area growth of thin films and heterointerfaces of such high $T_C$ layered ferromagnets. Substitution at the Fe sites in F5GT forming $(Fe_{1-x}M_x)_5GeTe_2$ (M = Ni, Co), leads to a significant enhancement of the magnetic ordering temperature. In Cobalt substituted samples, the $T_C$ increases up to ~360 K [25–28], while Ni substitution raises it further to the range of 450 to 492 K [29,30], However, the progress in the area of epitaxial and polycrystalline thin film growth has been limited to molecular beam epitaxy, with marginal success [2]. Further, unlike the case of vdW semiconductors such as WSe$_2$, the use of chemical vapor deposition techniques for high $T_C$ 2D magnets is challenging.

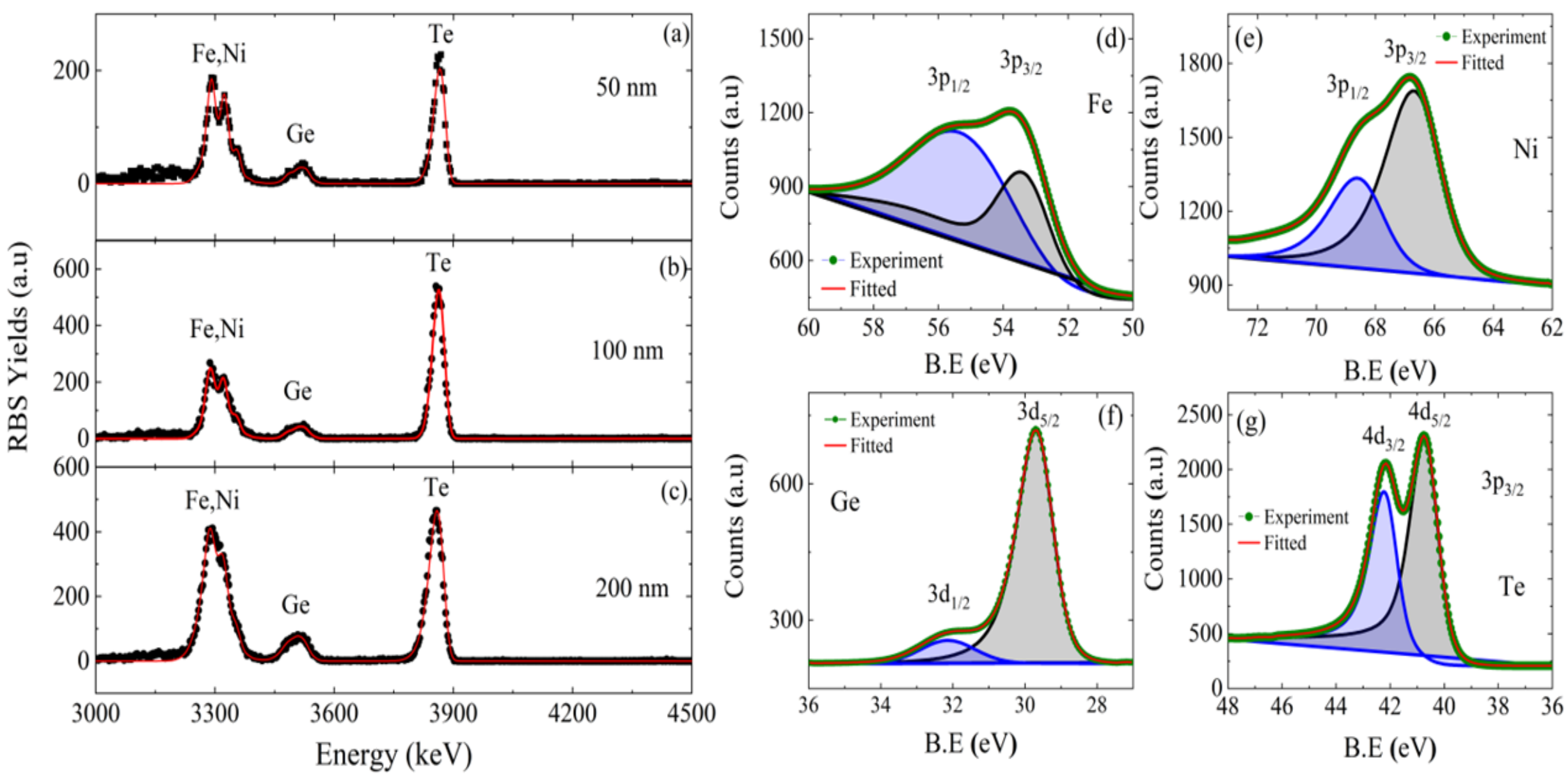

*Figure 1: RBS spectra of (Fe,Ni)$_5$GeTe$_2$ thin film of (a) 50 nm (b) 100 nm (c) 200 nm thickness. These spectra were collected with high energy (4.39 MeV) helium ions to resolve the backscattering yield of Ni and Fe, which is difficult with the low energy ions as the atomic masses of Fe and Ni are very close. X-ray photoelectron spectroscopy (XPS) analysis of the 100 nm (Fe,Ni)$_5$GeTe$_2$ thin film is presented in Figure 1(d–g), showing the elemental core-level spectra of (d) Fe, (e) Ni, (f) Ge (f), and (g) Te .*

Here we demonstrate the successful use of the pulsed laser deposition (PLD) technique to grow highly textured thin films of (Fe,Ni)$_5$GeTe$_2$ on the *c*-plane single crystals of sapphire. The films were characterized structurally using X-ray diffraction (XRD) which reveals a highly textured growth along the *c*-axis of the hexagonal unit cell. These vdW thin films whose stoichiometry has been established with Rutherford back scattering (RBS) and X-ray photoelectron spectroscopy (XPS), exhibit a robust ferromagnetic state up to ≈ 498 K, a notably high Curie temperature. Additionally, transport measurements reveal a distinct anomalous Hall effect (AHE) with Hall conductivity and Hall angle (%) of ~20 $\Omega^{-1}cm^{-1}$ and ~ 0.90 together with a pronounced thickness-dependent variation in magnetoresistance (MR).

Thin films of (Fe,Ni)$_5$GeTe$_2$ were deposited on single-side polished *c*-plane sapphire substrates using PLD with a KrF excimer laser ($\lambda$ = 248 nm). Structural and compositional analyses were performed using X-ray diffraction (Refer to the supplementary material for thin film deposition and XRD details), RBS and XPS. Electrical transport measurements, including Hall resistivity and magnetoresistance, were conducted on films patterned in a Hall bar geometry using a physical property measurement system equipped with a 9-tesla superconducting magnet.

Figure 1(a–c) presents the RBS spectra of (Fe,Ni)$_5$GeTe$_2$ thin films with nominal thicknesses of 50, 100, and 200 nm, deposited under identical conditions.

Measurements were performed using 4.39 MeV He ions, enabling clear resolution of Fe and Ni peaks, which is not feasible with lower energy beams due to their similar atomic masses. Elemental analysis yielded a composition of $(Fe+Ni)_{0.70}Ge_{0.08}Te_{0.22}$. The metal-to-Te and metal-to-Ge ratios were found to be ~3.0 and ~8.6, respectively—slightly exceeding the nominal values. The Fe/Ni ratio remained consistent, with a slight Ni enrichment across all thicknesses. Areal densities (in units of $10^{15}$ atoms/cm$^2$) of 69, 144, and 262 for the 50, 100, and 200 nm films, respectively, confirm thickness proportionality with a measured ratio of 1:2.1:3.8. Composition uncertainty analysis gives $(Fe+Ni)_{0.70\pm0.029}Ge_{0.080\pm0.003}Te_{0.22\pm0.010}$. Complementary XPS analysis of the 100 nm film [Fig. 1(d–g)], using a PHI Versa Probe II with a monochromatic Al $K_\alpha$ (hν = 1486.7 eV) source, confirmed the presence of Fe, Ni, Ge, and Te core-level peaks. Prior to acquisition, the sample was sputter-cleaned with 500 V $Ar^+$ ions to remove surface contamination. Spectra were recorded over a 200×200 μm$^2$ area with 0.05 eV steps and a 23 eV pass energy. Quantification from the Ge 3d, Te 4d, Fe 3p, and Ni 3p peaks using CASA XPS (v2.3.26) yielded a composition of $(Fe,Ni)_{0.62}Ge_{0.11}Te_{0.27}$, showing reasonable agreement with RBS values, with expected discrepancies due to XPS surface sensitivity.

The zero-field longitudinal resistivity ($\rho_{xx}$) of the $(Fe,Ni)_5GeTe_2$ films of three different thickness was measured over a temperature range of 10 to 400 K. Figure 2(a) shows the temperature-dependent resistivity ($\rho_{xx}$(T)), of 100 and 200 nm thick $(Fe,Ni)_5GeTe_2$ films, while the inset presents the data for the 50 nm film. The films exhibit metallic behavior. However, our measured resistivity is somewhat higher than the reported value for single crystals. However, Our measured resistivity is somewhat higher than the reported values on single crystals, which is typically in the range of 300–400 μΩ.cm [22,31,32]. To the best of our knowledge, there is no report on the resistivity values of $Fe_5GeTe_2$ thin films to compare with. The resistivity of the crystals is lower presumably because single crystals are generally defect-free, whereas the resistivity in textured films is affected by grain boundaries, defects, interface scattering and strain effects. It is worth noting that even in single crystals the room temperature resistivity is much higher than the Mooji's criterion [33,34] for the upper limit on the resistivity of 3D metal films, which is ≈ 150 μΩ.cm. This discrepancy may be attributed to the nature conduction in 2D materials, where the effective thickness of conducting channels may be much smaller that the thickness of the crystal that is used to calculate the resistivity.

To determine the $T_C$ of the $(Fe,Ni)_5GeTe_2$ films, the magnetization of the 200 nm thick film was measured at 0.5 tesla in the temperature range of 10 to 530 K using a high temperature vibrating sample magnetometer. As shown in Figure 2(b), the $T_C$ is estimated to be ≈ 498 K after subtracting the diamagnetic background of the substrate. The magnetization of 100 nm sample follows the same trend as of the 200 nm sample up to 400 K. However, the 50 nm sample could not be measured due to a very low signal. This enhancement in the ordering temperature is attributed to Ni substitution, which is consistent with previous reports on single crystals of $(Fe,Ni)_5GeTe_2$ [29,30]. The increase in $T_C$ is primarily driven by electron doping and structural modifications of F5GT on Ni substitution [29,30], which presumably alter the electronic band structure and density of states near the Fermi level, thereby modifying magnetic exchange interactions, likely through the Ruderman–Kittel–Kasuya–Yosida (RKKY) type mechanism. Structural insights further suggest that Ni preferentially substitutes at the Fe1 site, which is critical for establishing long-range ferromagnetic order due to its shortest Fe1–Fe3 bond length [29,30,35]. The inset of Fig.2(b) displays the isothermal magnetization curves measured at 300 K with the external magnetic field in the plane of the film and when it is normal to its surface. The saturation magnetization ($M_s$) is estimated to be ≈ 200 emu/cm³, which corresponds to ≈ 2.8 $\mu_B$ per formula unit. This value is consistent with previously reported results for Ni-doped $Fe_5GeTe_2$ crystals [30]. The anisotropy energy density calculated from the M(H) loops is ≈ 0.8 $J/cm^3$, which exceeds by 0.16-0.4 $J/cm^3$, the reported value for the pristine F5GT crystals [36,37], presumably due to the shape anisotropy of a thin film. The longitudinal and Hall resistivities, $\rho_{xx}$ and $\rho_{yx}$ of all three films were measured over a range of temperatures by cycling the out-of-plane (OOP) field from 0.0 to +5.0 to -5.0 to +5.0 tesla from a fully demagnetized state at a given temperature.

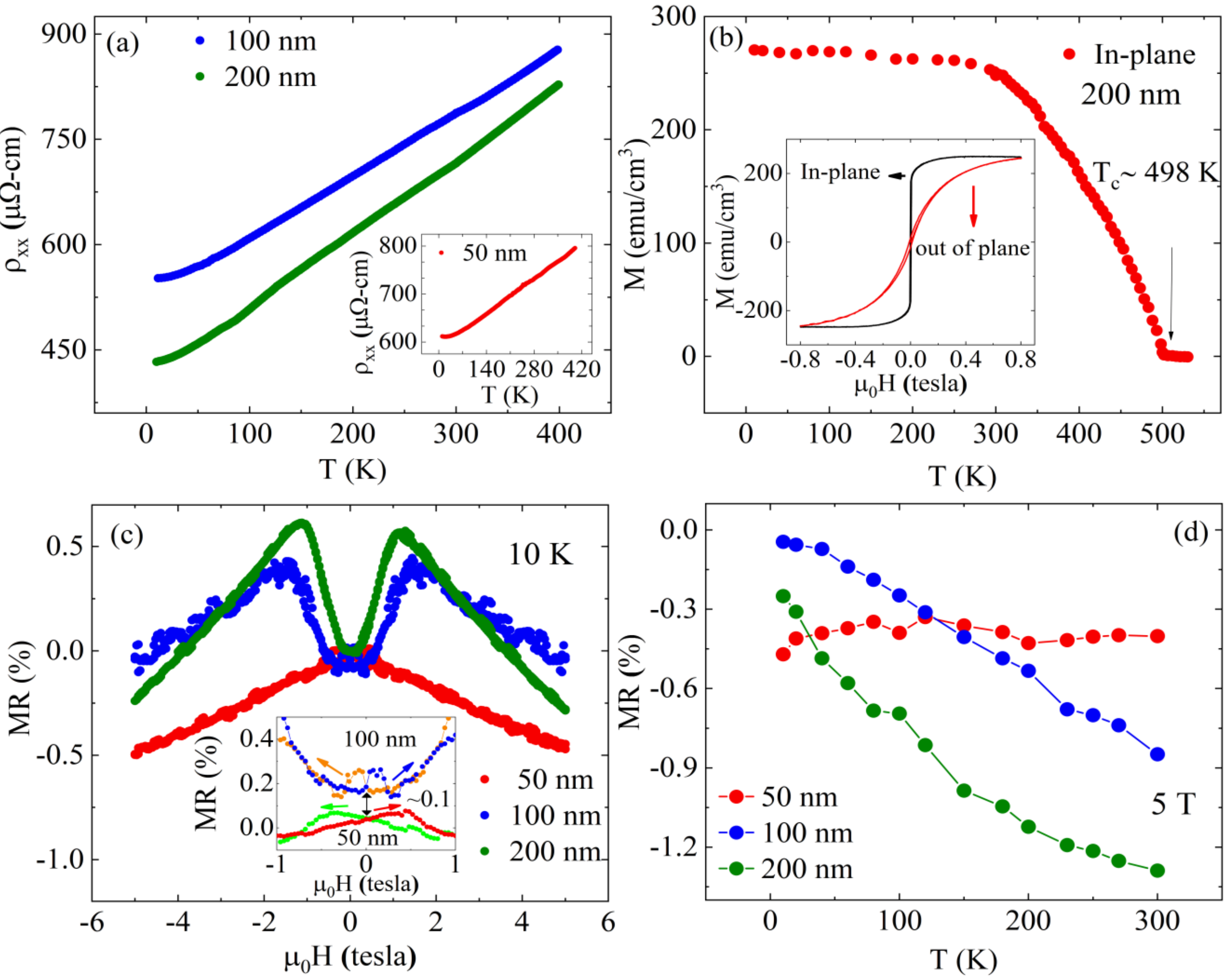

*Figure 2: (a) The temperature variation of the zero-field resistivity of the 100 and 200 nm thick films. Inset shows the resistivity of the 50 nm thick film. (b) The temperature dependence of the in-plane magnetization of the 200 nm thick film. The inset of Fig 2(b) shows field dependence of magnetization of the same film measured at 300 K in out-of-plane and in-plane field geometries. (c) MR (%) of the three films as a function of $\mu_0 H$ measured at 10 K. A magnified view of the cusp region, observed for both the 50 nm and 100 nm films, is shown in the inset of Fig. 2(c) (The blue and orange arrows indicate the MR (%) for the 100 nm film during increasing and decreasing magnetic field sweeps, respectively. Similarly, the red and green arrows represent the corresponding behavior for the 50 nm film). The MR data for the 100 nm film has been offset by 0.1 along the MR axis to clearly distinguish the two datasets in the inset. (d) Temperature variation of MR (%) of the films at 5 tesla.*

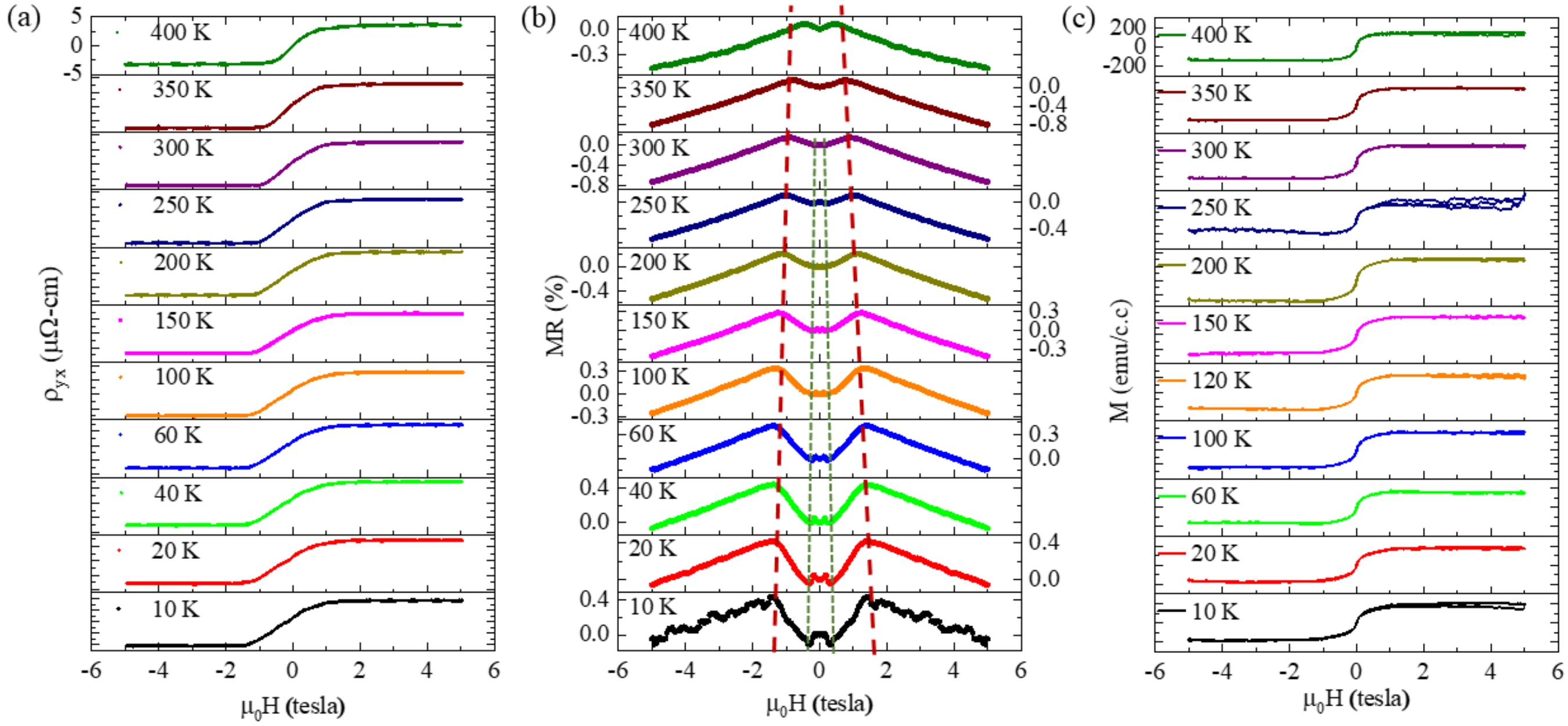

*Figure 3: (a) Magnetic field dependence of the Hall resistivity at different temperatures, from 10 to 400 K as the field is scanned from 0 to +5.0 to -5.0 and then back to +5.0 tesla. (b) The MR (%) in the same field and temperature range is shown in (b), and (c) displays the static magnetization loops measured as a function of magnetic field at the same set of temperatures as in (a).*

The magnetoresistance (MR) has been calculated using the standard relation:

$$MR(\%) = \left(\frac{\rho_{xx}(H)-\rho_{xx}(0)}{\rho_{xx}(0)}\right) \times 100 \qquad (1)$$

where, $\rho_{xx}(H)$ and $\rho_{xx}(0)$ denote the longitudinal resistivity in field H and at zero field, respectively.

The field dependence of the MR measured at 10 K and shown in Fig. 2(c) displays some interesting features which are discussed in combination with a magnified view of the MR curves shown in the inset of Fig. 2(c). Focusing on the inset first, the MR of the 50 nm thick film shows a distinct peak on crossing the zero-field line. A similar behavior is seen in the MR of the 100 nm film, although the position of these peaks is at a lower field. For the 200 nm film this feature is barely visible. These peaks are associate with the coercive field ($H_c$) of the magnetic film, [38] although the $H_c$ deduced from the M(H) is smaller. However, this difference can be understood in terms of the interaction of transport current with magnetic domain walls [38] in an MR measurement. A striking difference is seen in the overall field dependence of OOP MR of the 50 nm vs 100 and 200 nm films, as seen in the main panel of Fig. 2(c). While the MR of the 50 nm film is negative and increases linearly with field all the way to 9 tesla, for the 100 and 200 nm films it first increases to a critical field H* and then drops on increasing the field further. The value of H* compares well with the magnetization saturation field deduced from the M(H) loop measurements in the OOP geometry. This negative MR in the magnetically saturated state is ubiquitous in metallic ferromagnets and originated from *s-d* scattering [39]. We attribute the positive MR of the thicker films at H < H* to their complex domain structure whose reversal process may contribute to backscattering of charge carriers. A positive MR may also result from fluctuating domains which scatter electrons effectively before a stable domain structure evolves where the s-d scattering dominates the transport [40].

These features of the MR of the 100 nm thick film have been compared and discussed in detail with the behavior of $\rho_{yx}$ (H) and M (H) in the subsequent section. But first we present, in Fig. 2 (d), the temperature dependence of the MR (%) of all three samples at 5 tesla. A monotonic increase in the negative MR is seen with temperature in the case of the 100 and 200 nm thick films. This behavior is consistent with temperature dependence of MR seen in Heisenberg and double exchange ferromagnets as the $T_C$ is approached [41,42]. However, a distinctly different nature of MR (%) is seen in the case of the 50 nm thick film. Here it is small and remains constant on increasing the temperature.

Hall resistivity ($\rho_{yx}$) measurements were performed across a wide field–temperature (H-T) phase space to investigate the charge transport in the 100 nm-thick $(Fe,Ni)_5GeTe_2$ film. Figure 3(a) presents the $\rho_{yx}$ (H) data acquired at various temperatures ranging from 10 to 400 K. The Hall resistivity exhibits a nonlinear dependence on the applied magnetic field, tending towards saturation at fields above ≈ ± 1 tesla. The observed nonlinearity arises from the AHE, which coexists with the ordinary Hall effect. Analysis based on a single-band conduction model, which provides a reasonable approximation since the Hall resistivity remains linear in field above the saturation of the AHE, reveals hole-type carriers with a concentration of ≈ $7\times10^{21}$ $cm^{-3}$ and a mobility of ≈ 1.3$cm^2$/V-s at room temperature. Detailed temperature dependence is provided in Figure S2 of the Supplementary Information.

The total transverse resistivity measured in the standard Hall geometry can be expressed as:

$$\rho_{yx} = \rho_{yx}^{OHE} + \rho_{yx}^{AHE} \qquad (2)$$

Here, $\rho_{yx}^{OHE}$ corresponds to the ordinary Hall resistivity resulting from the Lorentz force acting on charge carriers and follows a linear field dependence as $R_0\mu_0H$, where $R_0$ is the Hall coefficient. The anomalous component, $\rho_{yx}^{AHE}$, originates from the magnetization of the sample and is expressed as $4\pi R_s M$, where $R_s$ is the anomalous Hall coefficient [43].

Figure 3(b) displays the magnetoresistance of the same 100 nm film across the temperature range of 10 to 400 K. At low fields (− 0.3 tesla ≤ $\mu_0$H ≤ 0.3 tesla), the MR exhibits an M-shaped profile at temperatures below ≈ 250 K (as shown by the green dotted line in Figure 3(b)). This behavior, previously described in the case of the 50 nm film, has its origin in magnetic coercivity of the samples. For a limited range of fields beyond the M shaped region, the MR is positive and increases with the field. This positive MR range, which lies between the red and green dotted lines in Fig. 3(b) is attributed to enhanced backscattering of charge carriers driven by the randomness in the alignment of domains prior to magnetic saturation [40]. For fields beyond the red dotted line, the MR is negative and increases monotonically with field. To correlate these features of MR with magnetization, we have measured the out-of-plane M(H) loops at several temperatures, as shown in Fig. 3(c). The magnetization approaches saturation at fields close to those corresponding to the red dotted line in Fig. 3(b).

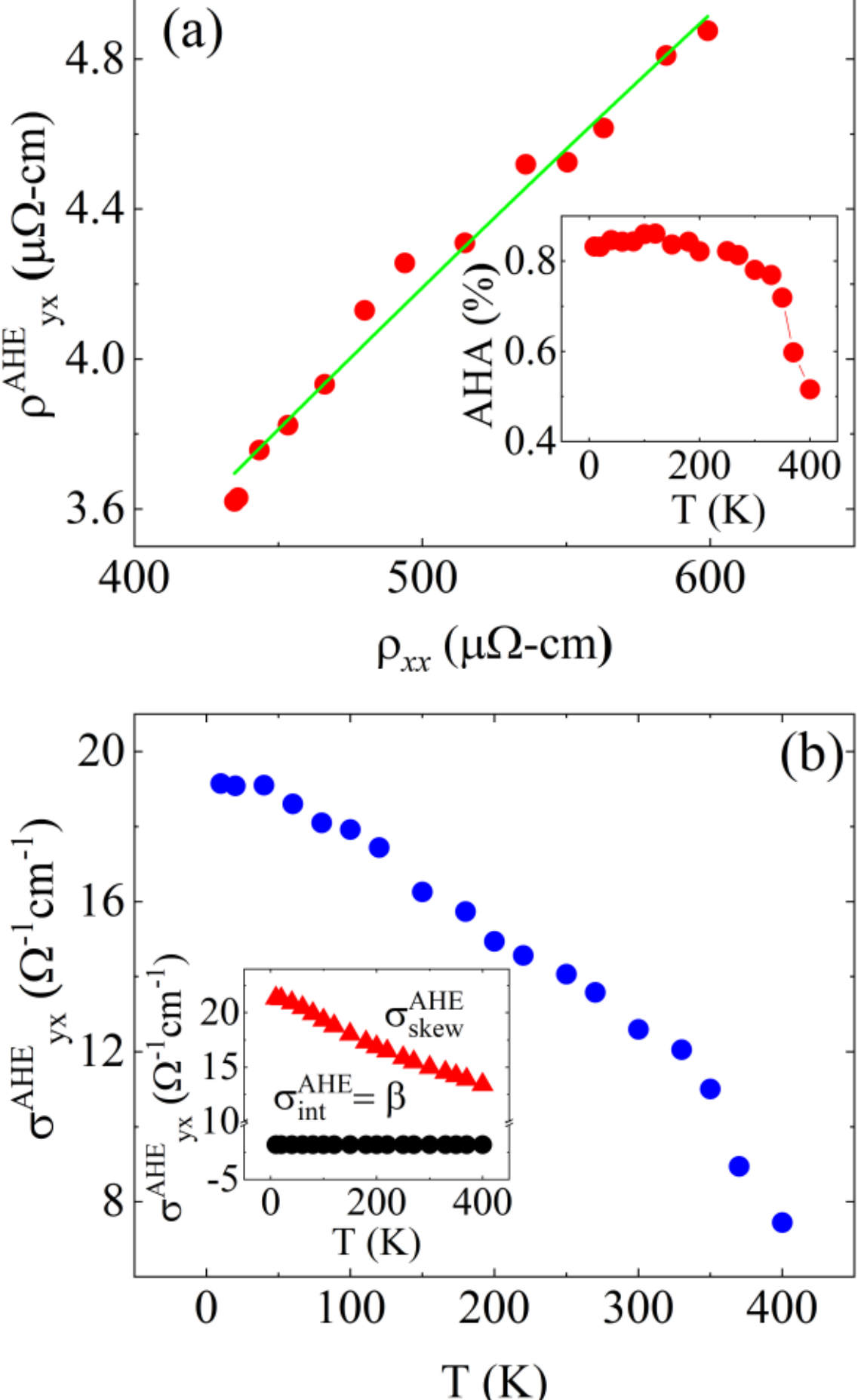

*Figure 4: (a) Fitting of anomalous Hall resistivity ($\rho_{xy}^{AHE}$) vs longitudinal resistivity ($\rho_{xx}$) using conventional scaling relation given in Eq. 3 (Green line) and the temperature dependence of AHA (inset). (b) The AHC as a function of temperature, and the variations of different contribution to AHC as a function of temperature (inset).*

The $\rho_{yx}^{AHE}$ at different temperatures has been calculated by subtracting the ordinary Hall contribution from the field-dependent $\rho_{yx}$. To quantitatively assess the contributions of intrinsic and extrinsic factors to the anomalous Hall resistivity, we fitted the $\rho_{yx}^{AHE}$ versus $\rho_{xx}$ data (red spheres in Fig. 4(a)) using the scaling relation [44,45]:

$$\rho_{yx}^{AHE} = \alpha\rho_{xx} + \beta\rho_{xx}^{2} \qquad (3)$$

where the coefficients α and β highlight the contributions of skew scattering and the combined effect of the electronic band structure and side jump scattering, respectively [43]. From this fitting, we obtained $\alpha \approx 0.0093$ and $\beta \approx -1.8$ $\mathrm{Ohm^{-1}cm^{-1}}$.

The anomalous Hall conductivity (AHC) $\sigma_{yx}^{AHE}$ has been calculated using the equation

$$\sigma_{yx}^{AHE} = -\rho_{yx}^{AHE}/({\rho_{yx}^{AHE}}^{2} + \rho_{xx}^{2}) \qquad (4)$$

and the result is shown in Fig. 4(b). The AHC at 10 K is ≈ 20 $\mathrm{Ohm^{-1}cm^{-1}}$ and drops with the increasing temperature. This value is lower than that reported for single crystals of F5GT [22,46] by a factor of ≈ 3.2. Using the values of $\alpha\rho_{xx}$ and β, the contributions of the intrinsic mechanism and skew scattering to the AHC have been estimated and the result is displayed in the inset of Fig. 4(b). Clearly, the AHE in $(Fe,Ni)_5GeTe_2$ is driven primarily by extrinsic skew scattering as seen in the figure. The anomalous Hall angle (AHA (%) $= (\frac{\sigma_{yx}^{AHE}}{\sigma_{xx}}) \times 100$) that quantifies the efficiency of longitudinal-to-transverse current conversion and plotted as a function of temperature in inset of Fig. 4(a) is relatively smaller by a factor of ≈ 2 compared to that of calculated F5GT single crystals [22,46]. These differences in AHC and AHA are suggestive of a key role of Ni substitution which decreases the overall magnetization and spin polarization of $Fe_5GeTe_2$, as Ni carries a smaller magnetic moment than Fe. Since skew scattering is a spin-dependent phenomenon, a reduction in spin polarization may suppress its contribution to AHE [43].

In conclusion, we have successfully demonstrated the synthesis of highly textured thin films of $(Fe,Ni)_5GeTe_2$ on *c*-plane sapphire using pulsed laser deposition. High resolution RBS and XPS techniques have been used to establish the elemental concentration in the films, which exhibit robust ferromagnetic ordering with a $T_C$ reaching ≈ 498 K. Electrical transport measurements show AHE, with AHC of ≈ 20 $Ohm^{-1}cm^{-1}$ arising primarily from the extrinsic scattering mechanisms leading to a Hall angle (%) of ≈ 0.90. A comparison of these AHE parameters with those for F5GT crystals establishing the role of Nickel substitution. Additionally, the MR of these metallic films displays a pronounced dependence on their thickness and the strength of applied magnetic field, highlighting the tunability of spin-dependent transport in this high $T_C$ vdW ferromagnet.

**ACKNOWLEDGMENT:**

This research has been performed at the United States Department of Defense funded Center of Excellence for Advanced Electro-photonics with 2D materials ‑ Morgan State University, under grant No. W911NF2120213. The authors also thank to Prof. Abdellah Lisfi for his assistance with the high temperature VSM measurements.